\documentclass[12pt,a4paper]{article}

\newcommand{\np}{\mbox{NP}}
\newcommand{\p}{\mbox{P}}
\newcommand{\uf}{\mbox{UF}}
\newtheorem{lemma}{Lemma}{\bf }{\it }
\newtheorem{theorem}{Theorem}{\bf }{\it }
\newtheorem{corollary}{Corollary}{\bf }{\it }

\title{A class of problems of \np\ to be worth to search
an efficient solving algorithm}

\vspace{24pt}

\author{Anatoly D. Plotnikov\thanks{
Vinnitsa Institute of Regional Economics and Management,
e-mail: aplot@tom.vinnica.ua}}
\date{}

\begin{document}
\maketitle

\begin{abstract}
We examine possibility to design an efficient solving algorithm for problems
of the class \np. It is introduced a classification of \np\ problems by the 
property that a partial solution of size $k$ can be extended into a partial 
solution of size $k+1$ in polynomial time. It is defined an unique class 
problems to be worth to search an efficient solving algorithm. The problems,
which are outside of this class, are inherently exponential.
We show that the Hamiltonian cycle problem is inherently exponential.
\end{abstract}

\section{Introduction}
\label{IN}

A problem $Z$ belongs to the class \np\ if it has a finite input of size $n$,
a finite output (a solution) of size $p(n)$, where $p(n)$ is some polynomial, 
and verifying time of the solution is also a polynomial on $n$.

Every problem of \np\ is solvable in classical sense since it
can be solved by a deterministic Turing machine \cite{gary-johnson}.

A solving algorithm of a problem $Z\in \np$ is called {\it efficient} if the 
solution of $Z$ can be obtained in the polynomial number of steps on $n$. 
A set of problems of \np, having a polynomial solving algorithm, is denoted 
by \p. 

One of the main achievements in designing of solving algorithms is joined 
with Matroid Theory.

Let $R$ be some finite set, and $Q$ be a non-empty set of subsets of $R$.
A two ($R, Q$), satisfying property 
\begin{equation}
\label{1}
\mbox{if}\ \pi_{1}\in Q\ \mbox{and}\ \pi_{2}\subset \pi_{1}\ 
\mbox{then}\ \pi_{2}\in Q
\end{equation}
is called a {\it hereditary} system.

A hereditary system $M$ = ($R, Q$) is called a {\it matroid} when it 
satisfies the following property \cite{wilson}:
\begin{equation}
\label{2}
\begin{array}{l}
\mbox{if}\ \pi_{1}, \pi_{2}\in Q,\ \mbox{and}\ 
Card(\pi_{2})=Card(\pi_{1})+1\ \mbox{then there exists}\\ \mbox{an element}\ 
r\in \pi_{2}\setminus \pi_{1}\ \mbox{such that}\ 
\pi_{1}\cup \{r\}\in Q,
\end{array}
\end{equation}
where $r\in R$.

If an optimization problem of \np\ satisfies properties (\ref{1}) and 
(\ref{2}) then it can be solved by means of a {\it greedy} algorithm. A greedy 
algorithm make locally optimal choices at each step in the hope that these 
choices will produce a globally optimal solution \cite{baase, 
swamy-thulasiraman, west}.

The following assertion takes place \cite{swamy-thulasiraman}.

\begin{theorem}
\label{greedy}
Let ($R, Q$) be a hereditary system. Then a greedy algorithm produces
an maximum element of $Q$ if and only if ($R, Q$) be a matroid.
\end{theorem}

Unfortunately, there are problems of \np\ for which an efficient 
solving algorithm is unknown. An analysis of \np\ problems shows that the
main difficulty is exhaustive enumeration of the elements of the solution.

The following questions raise: What are causes for appearing of exhaustive 
enumeration? When can we design an efficient solving algorithm?

In this paper we show that one need to analyze a mathematical model of a
problem. We define a class problem of \np\  be worth to search an efficient 
solving algorithm. We show also that other problems of \np\ are inherently 
exponential. Consequently, for similar problems the finding of the efficient 
solving algorithm is senseless.

\section{Mathematical model}
\label{MOD}

Let ($R, Q$) be a hereditary system. Further let $P$ be a predicate 
system. For each subset $R_{j}$ of $R$ this system allows to find a value 
of a predicate ``$R_{j}\in Q$?'' Let it is required to find a subset 
$R_{j}\subset R$ such that $R_{j}\in Q$.

Many important problems of \np\ have the similar statement.

\vspace{1pc}

\noindent
{\bf Satisfiability Problem (SAT).} Let $\phi$ be a Boolean expression over 
$n$ variables $x_{1}, \ldots , x_{n}$ in conjunctive normal form. It is
required to find values of the variables which make $\phi$ equal ``true'',
that is, co-called truth assignment of variables. Let $R$ be a set of literals 
$r$, where $r$ is either $x_{i}$ or ${\bar x}_{i}$ ($i=\overline {1, n}$). 
The literals $x$ and ${\bar x}$ we shall call {\it contrary}.

Further, let $\cal B$ be a set of subsets $\pi^{*}$ of $R$ determining truth 
assignments of variables. Denote by a set $Q$ of subsets $\pi$ of $R$ such 
that $\pi\in Q$ if and only if $\pi\subseteq \pi^{*}$, 
$\pi^{*}\in {\cal B}$. Obviously that a two ($R, Q$) is a hereditary 
system.

\vspace{1pc}

\noindent
{\bf Hamiltonian Cycle Problem (HCP).} Let $G=(V, E)$ be a $n$-vertex 
undirected graph. It is required to find a cycle of edges in $G$ which includes 
each of the $n$ vertices exactly once. Let $\cal B$ be a set of Hamiltonian
cycles $\pi^{*}$ of $G$. Denote by a set $Q$ of subsets $\pi$ of $E$ such 
that $\pi\in Q$ if and only if $\pi\subseteq \pi^{*}$, 
$\pi^{*}\in {\cal B}$. It is clear that a two ($E, Q$) is a hereditary 
system.

\vspace{1pc}

Consider a hereditary system ($R, Q$).

Let $w(r_{i})$ ($i=\overline {1, n}$) be an integer, is called by a {\it weight} 
of the element $r_{i}$ of $R$. For every $\pi\in Q$ we define a sum
\begin{displaymath}
w(\pi)=\sum_{\forall r\in \pi} w(r).
\end{displaymath}
This sum we shall call a {\it weight} of $\pi$.

Let it is required to find an element $\pi^{*}$ of $Q$, having the maximum 
weight.

We have an optimization problem.

\vspace{1pc}

\noindent
{\bf Maximum Independent Set Problems (MISP).} Let $G=(V, E)$ be a $n$-vertex 
undirected graph. It is required to find a subset $\pi^{*}\subseteq V$, 
having the maximum number vertices, such that every two vertices in $\pi^{*}$
are non-adjacent in $G$. A set $\pi\subseteq V$ is called independent if every
two vertices in $\pi$ are non-adjacent. Let $Q$ be a set of all
independent sets of vertices in of $G$. It is easy to see that a two 
($V, Q$) is a hereditary system. In this problem $w(\pi)=Card(\pi)$
for any $\pi\in Q$.

\vspace{1pc}

Let there be a problem $Z\in \np$ which is a hereditary system ($R, Q$).
Any element $\pi\in Q$ we shall call an {\it admissible} solution
of $Z$. An inclusion maximal admissible solution $\pi^{*}$ of $Z$ is called 
{\it support}. An admissible solution $\pi$ of $Z$ is called {\it partial} if 
there exists a support solution $\pi^{*}$ such that $\pi\subseteq \pi^{*}$.
If $\pi$ be some support solution of the problem $Z$, and $\pi_{1}$ be some 
partial solution of this problem such that $\pi_{1} \subset \pi $ then the 
partial solution $\pi_{1}$ will be {\it own} for support solution $\pi$.

We considered a {\it computational} model of the problems in \np. In 
Complexity Theory, every problem of \np\ is considered as a {\it decision} 
problem. A decision problem is a computational problem whose solution is 
{\it yes} or {\it no} \cite{loui}. The solution of a computational problem 
(in given case it is some of support solutions) we may consider as ``proof'' 
that the corresponding decision problem of \np\ has an answer ``yes''. 
Therefore, the conception of admissible solution is more wide than the 
conception of ``proof'' for a decision problem. 

\section{Sequential method}
\label{SE}

Let there be a problem $Z\in \np$. We assume that $Z$ is a hereditary system
($R, Q$). The questions raises: {\it How} can we construct an admissible 
solution of $Z$?

A Turing machine is a generally accepted model of computation (see, for 
example, \cite{aho, loui}). Therefore, we may think that we have an 
1-type Turing machine $M_{1}$. The machine $M_{1}$ produces symbols into cells 
of tape {\it sequentially}, that is, symbol-by-symbol. If we shall be believe
that Turing machine solve the problem $Z$ then we may consider a result of 
such work of $M_{1}$ at each step as an admissible solution of $Z$. Naturally 
to consider the record of a symbol on tape as construction of the next element 
of admissible solution.

Thus, procedure of construction of admissible solution $\pi\in Q$ is extended 
on the time, i.e. elements of one are obtained element-by-element.

A method of the construct of the required solution, when we are obtaining
its elements by step-by-step, element-by-element, will be called 
{\it sequential}.

Let $\pi_{1}$, $\pi$ be respectively the partial and support solutions of a 
problem $Z \in \np$ such that $\pi_{1} \subset \pi$. Denote the construction 
time for these partial and support solutions of $Z$ by $t(\pi_{1})$ and 
$t(\pi)$ respectively. Then the following assertion takes place.

\begin{theorem}
\label{time}
$t(\pi_{1}) < t(\pi)$.
\end{theorem}

\noindent
{\bf Proof.} By the definition of the sequential method, every of support 
solutions can be obtained {\it after} constructing of an own partial solution. 
That proves Theorem \ref{time}.$\circ$

\begin{theorem}
\label{seq}
The solution of any problem $Z \in \np$ can be obtained by a sequential 
method.
\end{theorem}

\noindent
{\bf Proof.} We believe that each problem of the class \np\ is solvable (see 
Section \ref{IN}), that is, each of such problem can be solved by the 
deterministic Turing machine. Since this machine works sequentially, it 
produces the solution of a problem by step-by-step, element-by-element. 
Therefore Theorem \ref{seq} is true.$\circ$

\vspace{1pc}

Obviously, one can believe that a sequential method is a sole {\it general} 
method of the solving for every problem $Z \in \np$.

In fact, for example, let there is necessary to find some independent set
of the graph vertices. Obviously, in common case the {\it simultaneous choice} 
of a number of such vertices is impossible if the graph structure was unknown 
beforehand. Clearly, each subsequent vertices can be chosen only if it is known 
which vertices was chosen in the formed independent set before.

\section{Problems without lookahead}

Let $\pi_{1}$ be a partial solution of the problem $Z\in \np$. By designing of 
the next partial solution, the problems of \np\ can be partitioned into two 
classes \cite{plot:turing}:

\begin{itemize}
\item the problems for which the next partial solution 
$\pi_{2}=\pi_{1}\cup \{r\}$ can be found in polynomial time on early found 
partial solution by picking one element of the set $R\setminus \pi_{1}$;
\item all other problems of \np.
\end{itemize}

That is, the problems of \np\ can be classified depending on the computing
time of the predicate ``$\pi_{1} \cup \{r\}\in Q$?'' for every partial solution 
$\pi_{1} \in Q$ and for any element $r \in R \setminus \pi_{1}$. If such 
predicate can be computed in polynomial time on the problem dimension then 
such problem will be called {\it the problem without lookahead}. Otherwise 
the problem is called {\it inherently exponential}.

A set of all problems without lookahead we will denote by \uf, where 
$\uf\subseteq \np$. 

\begin{theorem}
\label{support}
\label{a} A support solution of a problem $Z\in \np$ can be found in
polynomial time if and only if $Z\in \uf$.
\end{theorem} 

\noindent
{\bf Proof.} Let there be the problem $Z \in \np$ such that $Z \in \uf$. By 
definition of problems without lookahead, the next partial solution of $Z$ 
can be found in polynomial time. Since $\oslash\in Q$ for any $Z\in \uf$, and 
a support solution contains at the most $n$ elements, where $n$ is a problem 
size, then it implies polynomial construction time of the support solution.

On the other hand, let there be the problem $Z \in \np$ such which is resolved 
in polynomial time. Suppose that $Z \not\in \uf$. In this case there exists 
at least one of the partial solutions of the problem $Z$, determined in 
exponential time. By the condition of Theorem \ref{support}, the support 
solution of $Z$ is found in polynomial time. We have given the contradiction 
of the Theorems \ref{time} and \ref{seq}.$\circ$

\vspace{1pc}

Thus, the class \uf\ is induced by problems of \np\ for which a support
solution may be construct in polynomial time. Notice that this support 
solution may not be the global solution of the problem.

Prove that the HCP is outside of the class \uf.

Consider the following optimization problem. 

Let $G=(X, E)$ be an undirected graph without loops and multiple edges, where 
$X$ is the vertex set of $G$, and $E$ is the edge set. Simple cycles $C_{i}$,
$C_{p}$ ($i\not= p$) of $G$ are called {\it disjoint} if they have no common 
vertices. A collection $\pi$ = $\{C_{1}, \ldots , C_{k}$ is called a
{\it partition of $G$ into disjoint edges and/or cycles} if

\begin{itemize}
\item 
for each pair of cycles $C_{i}, C_{p}\in \pi$ ($i\not= p$) 
$X(C_{i})\cap X(C_{p})=\oslash$;
\item 
and
\[
\bigcup_{\forall C_{i}\in \pi} X(C_{i}) = X,
\]
\end{itemize}
where $X(C_{i})$ is a set of vertices belonging to the cycle (or the edge) 
$C_{i}$.

It is required to find a partition $\pi^{*}$ having the minimum number of 
cycles (edges).

This problem is \np-complete, and it is formulated as the Minimum Vertex 
Disjoint Cycle Cover Problem (MVDCCP).

It is known that an admissible solution of MVDCCP can be obtained as a solution
of the assignment problem (see, for example, \cite{christofides, lovasz})
in polynomial time. Hence, MVDCCP belongs to \uf. Its support solution -- 
some partition of $G$ into disjoint edges and/or cycles -- may be constructed
in polynomial time.

On the other hand, it is evident MVDCCP is not a matroid. Therefore, locally 
optimal choice does not guarantee that the obtained solution of the problem 
will be optimal. 

\begin{figure}[htbp]
\begin{center}
\unitlength 1.00mm
\linethickness{0.4pt}
\begin{picture}(112.17,113.26)
\put(12.27,10.11){\line(0,1){40.00}}
\put(12.27,45.11){\line(1,0){40.00}}
\put(52.27,50.11){\line(0,-1){40.00}}
\put(52.27,10.11){\line(-1,0){40.00}}
\put(12.27,50.11){\line(1,0){40.00}}
\put(12.27,45.11){\line(1,0){40.00}}
\put(12.27,40.11){\line(1,0){40.00}}
\put(12.27,35.11){\line(1,0){40.00}}
\put(12.27,30.11){\line(1,0){40.00}}
\put(12.27,25.11){\line(1,0){40.00}}
\put(12.27,20.11){\line(1,0){40.00}}
\put(12.27,15.11){\line(1,0){40.00}}
\put(17.27,50.11){\line(0,-1){40.00}}
\put(22.27,50.11){\line(0,-1){40.00}}
\put(27.27,50.11){\line(0,-1){40.00}}
\put(32.27,50.11){\line(0,-1){40.00}}
\put(37.27,50.11){\line(0,-1){40.00}}
\put(42.27,50.11){\line(0,-1){40.00}}
\put(47.27,50.11){\line(0,-1){40.00}}
\put(19.77,47.69){\makebox(0,0)[cc]{1}}
\put(49.77,47.69){\makebox(0,0)[cc]{1}}
\put(49.77,47.71){\circle{3.90}}
\put(14.77,42.69){\makebox(0,0)[cc]{1}}
\put(14.77,42.71){\circle{3.90}}
\put(24.77,42.69){\makebox(0,0)[cc]{1}}
\put(34.77,42.69){\makebox(0,0)[cc]{1}}
\put(49.77,42.69){\makebox(0,0)[cc]{1}}
\put(19.77,37.69){\makebox(0,0)[cc]{1}}
\put(29.77,37.69){\makebox(0,0)[cc]{1}}
\put(39.77,37.69){\makebox(0,0)[cc]{1}}
\put(39.77,37.71){\circle{3.90}}
\put(24.77,32.69){\makebox(0,0)[cc]{1}}
\put(24.77,32.71){\circle{3.90}}
\put(34.77,32.69){\makebox(0,0)[cc]{1}}
\put(39.77,32.69){\makebox(0,0)[cc]{1}}
\put(19.77,27.69){\makebox(0,0)[cc]{1}}
\put(29.77,27.69){\makebox(0,0)[cc]{1}}
\put(44.77,27.69){\makebox(0,0)[cc]{1}}
\put(44.77,27.71){\circle{3.90}}
\put(24.77,22.69){\makebox(0,0)[cc]{1}}
\put(29.77,22.69){\makebox(0,0)[cc]{1}}
\put(29.77,22.71){\circle{3.90}}
\put(44.77,22.69){\makebox(0,0)[cc]{1}}
\put(34.77,17.69){\makebox(0,0)[cc]{1}}
\put(34.77,17.71){\circle{3.90}}
\put(39.77,17.69){\makebox(0,0)[cc]{1}}
\put(49.77,17.69){\makebox(0,0)[cc]{1}}
\put(14.77,12.69){\makebox(0,0)[cc]{1}}
\put(19.77,12.69){\makebox(0,0)[cc]{1}}
\put(19.77,12.71){\circle{3.90}}
\put(44.77,12.69){\makebox(0,0)[cc]{1}}
\put(9.27,47.61){\makebox(0,0)[cc]{$x_{1}$}}
\put(9.27,42.61){\makebox(0,0)[cc]{$x_{2}$}}
\put(9.27,37.61){\makebox(0,0)[cc]{$x_{3}$}}
\put(9.27,32.61){\makebox(0,0)[cc]{$x_{4}$}}
\put(9.27,27.61){\makebox(0,0)[cc]{$x_{5}$}}
\put(9.27,22.61){\makebox(0,0)[cc]{$x_{6}$}}
\put(9.27,17.61){\makebox(0,0)[cc]{$x_{7}$}}
\put(9.27,12.61){\makebox(0,0)[cc]{$x_{8}$}}
\put(14.27,53.11){\makebox(0,0)[cc]{$x_{1}$}}
\put(19.27,53.11){\makebox(0,0)[cc]{$x_{2}$}}
\put(24.34,53.11){\makebox(0,0)[cc]{$x_{3}$}}
\put(29.27,53.11){\makebox(0,0)[cc]{$x_{4}$}}
\put(34.34,53.11){\makebox(0,0)[cc]{$x_{5}$}}
\put(39.34,53.11){\makebox(0,0)[cc]{$x_{6}$}}
\put(44.27,53.11){\makebox(0,0)[cc]{$x_{7}$}}
\put(49.27,53.11){\makebox(0,0)[cc]{$x_{8}$}}
\put(62.49,10.11){\line(0,1){40.00}}
\put(62.49,45.11){\line(1,0){40.00}}
\put(102.49,50.11){\line(0,-1){40.00}}
\put(102.49,10.11){\line(-1,0){40.00}}
\put(62.49,50.11){\line(1,0){40.00}}
\put(62.49,45.11){\line(1,0){40.00}}
\put(62.49,40.11){\line(1,0){40.00}}
\put(62.49,35.11){\line(1,0){40.00}}
\put(62.49,30.11){\line(1,0){40.00}}
\put(62.49,25.11){\line(1,0){40.00}}
\put(62.49,20.11){\line(1,0){40.00}}
\put(62.49,15.11){\line(1,0){40.00}}
\put(67.49,50.11){\line(0,-1){40.00}}
\put(72.49,50.11){\line(0,-1){40.00}}
\put(77.49,50.11){\line(0,-1){40.00}}
\put(82.49,50.11){\line(0,-1){40.00}}
\put(87.49,50.11){\line(0,-1){40.00}}
\put(92.49,50.11){\line(0,-1){40.00}}
\put(97.49,50.11){\line(0,-1){40.00}}
\put(69.99,47.69){\makebox(0,0)[cc]{1}}
\put(69.99,47.71){\circle{3.90}}
\put(99.99,47.69){\makebox(0,0)[cc]{1}}
\put(64.99,42.69){\makebox(0,0)[cc]{1}}
\put(74.99,42.69){\makebox(0,0)[cc]{1}}
\put(74.99,42.71){\circle{3.90}}
\put(84.99,42.69){\makebox(0,0)[cc]{1}}
\put(99.99,42.69){\makebox(0,0)[cc]{1}}
\put(69.99,37.69){\makebox(0,0)[cc]{1}}
\put(79.99,37.69){\makebox(0,0)[cc]{1}}
\put(89.99,37.69){\makebox(0,0)[cc]{1}}
\put(89.99,37.71){\circle{3.90}}
\put(74.99,32.69){\makebox(0,0)[cc]{1}}
\put(84.99,32.69){\makebox(0,0)[cc]{1}}
\put(84.99,32.71){\circle{3.90}}
\put(89.99,32.69){\makebox(0,0)[cc]{1}}
\put(69.99,27.69){\makebox(0,0)[cc]{1}}
\put(79.99,27.69){\makebox(0,0)[cc]{1}}
\put(94.99,27.69){\makebox(0,0)[cc]{1}}
\put(94.99,27.71){\circle{3.90}}
\put(74.99,22.69){\makebox(0,0)[cc]{1}}
\put(79.99,22.71){\circle{3.90}}
\put(79.99,22.69){\makebox(0,0)[cc]{1}}
\put(94.77,22.69){\makebox(0,0)[cc]{1}}
\put(84.99,17.69){\makebox(0,0)[cc]{1}}
\put(89.99,17.69){\makebox(0,0)[cc]{1}}
\put(99.77,17.69){\makebox(0,0)[cc]{1}}
\put(99.99,17.71){\circle{3.90}}
\put(64.99,12.69){\makebox(0,0)[cc]{1}}
\put(64.99,12.71){\circle{3.90}}
\put(69.99,12.69){\makebox(0,0)[cc]{1}}
\put(94.99,12.69){\makebox(0,0)[cc]{1}}
\put(59.49,47.61){\makebox(0,0)[cc]{$x_{1}$}}
\put(59.49,42.61){\makebox(0,0)[cc]{$x_{2}$}}
\put(59.49,37.61){\makebox(0,0)[cc]{$x_{3}$}}
\put(59.49,32.61){\makebox(0,0)[cc]{$x_{4}$}}
\put(59.49,27.61){\makebox(0,0)[cc]{$x_{5}$}}
\put(59.49,22.61){\makebox(0,0)[cc]{$x_{6}$}}
\put(59.49,17.61){\makebox(0,0)[cc]{$x_{7}$}}
\put(59.49,12.61){\makebox(0,0)[cc]{$x_{8}$}}
\put(64.49,53.11){\makebox(0,0)[cc]{$x_{1}$}}
\put(69.49,53.11){\makebox(0,0)[cc]{$x_{2}$}}
\put(74.56,53.11){\makebox(0,0)[cc]{$x_{3}$}}
\put(79.49,53.11){\makebox(0,0)[cc]{$x_{4}$}}
\put(84.56,53.11){\makebox(0,0)[cc]{$x_{5}$}}
\put(89.56,53.11){\makebox(0,0)[cc]{$x_{6}$}}
\put(94.49,53.11){\makebox(0,0)[cc]{$x_{7}$}}
\put(99.49,53.11){\makebox(0,0)[cc]{$x_{8}$}}
\put(29.49,3.56){\makebox(0,0)[cc]{(c)}}
\put(79.71,3.78){\makebox(0,0)[cc]{(d)}}
\put(15.00,69.70){\line(1,0){20.00}}
\put(35.00,69.70){\line(0,1){40.00}}
\put(15.00,99.70){\line(1,0){20.00}}
\put(15.00,99.70){\circle*{3.00}}
\put(5.00,84.70){\circle*{3.00}}
\put(15.00,69.70){\circle*{3.00}}
\put(15.00,109.70){\circle*{3.00}}
\put(35.00,109.70){\circle*{3.00}}
\put(35.00,99.70){\circle*{3.00}}
\put(50.00,79.70){\circle*{3.00}}
\put(35.00,69.70){\circle*{3.00}}
\put(1.78,89.70){\makebox(0,0)[cc]{$x_{1}$}}
\put(8.89,101.03){\makebox(0,0)[cc]{$x_{2}$}}
\put(8.67,112.59){\makebox(0,0)[cc]{$x_{3}$}}
\put(39.33,113.26){\makebox(0,0)[cc]{$x_{4}$}}
\put(38.44,95.26){\makebox(0,0)[cc]{$x_{5}$}}
\put(49.78,74.14){\makebox(0,0)[cc]{$x_{6}$}}
\put(36.00,65.26){\makebox(0,0)[cc]{$x_{7}$}}
\put(15.56,65.26){\makebox(0,0)[cc]{$x_{8}$}}
\put(5.00,84.70){\line(2,3){9.93}}
\put(5.15,84.55){\line(2,3){9.93}}
\put(5.30,84.40){\line(2,3){9.93}}
\put(4.85,84.85){\line(2,3){9.93}}
\put(4.70,85.00){\line(2,3){9.93}}
\put(15.00,99.70){\line(0,1){10.00}}
\put(15.00,109.70){\line(6,-5){35.00}}
\put(15.15,109.85){\line(6,-5){35.00}}
\put(15.30,110.00){\line(6,-5){35.00}}
\put(14.85,109.55){\line(6,-5){35.00}}
\put(14.70,109.40){\line(6,-5){35.00}}
\put(50.00,80.38){\line(-1,2){15.00}}
\put(50.15,80.53){\line(-1,2){15.00}}
\put(50.30,80.68){\line(-1,2){15.00}}
\put(49.85,80.23){\line(-1,2){15.00}}
\put(49.70,80.08){\line(-1,2){15.00}}
\put(15.00,69.70){\line(-2,3){10.00}}
\put(15.15,69.85){\line(-2,3){10.00}}
\put(15.30,70.00){\line(-2,3){10.00}}
\put(14.85,69.55){\line(-2,3){10.00}}
\put(14.70,69.40){\line(-2,3){10.00}}
\put(75.67,99.70){\line(0,-1){30.00}}
\put(95.67,109.70){\line(-1,0){20.00}}
\put(75.67,99.70){\line(1,0){20.00}}
\put(75.67,99.70){\circle*{3.00}}
\put(65.67,84.70){\circle*{3.00}}
\put(75.67,69.70){\circle*{3.00}}
\put(75.67,109.70){\circle*{3.00}}
\put(95.67,109.70){\circle*{3.00}}
\put(95.67,99.70){\circle*{3.00}}
\put(110.67,79.70){\circle*{3.00}}
\put(95.67,69.70){\circle*{3.00}}
\put(62.45,89.70){\makebox(0,0)[cc]{$x_{1}$}}
\put(69.56,101.03){\makebox(0,0)[cc]{$x_{2}$}}
\put(69.34,112.59){\makebox(0,0)[cc]{$x_{3}$}}
\put(100.00,113.26){\makebox(0,0)[cc]{$x_{4}$}}
\put(99.11,95.26){\makebox(0,0)[cc]{$x_{5}$}}
\put(110.45,74.14){\makebox(0,0)[cc]{$x_{6}$}}
\put(96.67,65.26){\makebox(0,0)[cc]{$x_{7}$}}
\put(76.23,65.26){\makebox(0,0)[cc]{$x_{8}$}}
\put(65.82,84.55){\line(2,3){9.93}}
\put(65.97,84.40){\line(2,3){9.93}}
\put(65.67,84.70){\line(2,3){9.93}}
\put(65.52,84.85){\line(2,3){9.93}}
\put(65.37,85.00){\line(2,3){9.93}}
\put(75.67,109.70){\line(6,-5){35.00}}
\put(75.82,109.85){\line(6,-5){35.00}}
\put(75.97,110.00){\line(6,-5){35.00}}
\put(75.52,109.55){\line(6,-5){35.00}}
\put(75.37,109.40){\line(6,-5){35.00}}
\put(110.67,80.38){\line(-1,2){15.00}}
\put(110.82,80.53){\line(-1,2){15.00}}
\put(110.97,80.68){\line(-1,2){15.00}}
\put(110.52,80.23){\line(-1,2){15.00}}
\put(110.37,80.08){\line(-1,2){15.00}}
\put(75.67,69.70){\line(-2,3){10.00}}
\put(75.82,69.85){\line(-2,3){10.00}}
\put(75.97,70.00){\line(-2,3){10.00}}
\put(75.52,69.55){\line(-2,3){10.00}}
\put(75.37,69.40){\line(-2,3){10.00}}
\put(35.15,69.70){\line(3,2){15.15}}
\put(95.76,69.70){\line(3,2){15.15}}
\put(15.11,69.78){\line(1,0){20.00}}
\put(25.15,61.52){\makebox(0,0)[cc]{(a)}}
\put(85.15,61.82){\makebox(0,0)[cc]{(b)}}
\linethickness{3.0pt}
\put(75.67,99.70){\line(0,1){10.00}}
\put(15.00,99.70){\line(0,-1){30.00}}
\put(35.00,69.70){\line(0,1){30.00}}
\put(95.67,69.70){\line(0,1){40.00}}
\put(35.00,109.70){\line(-1,0){20.00}}
\put(75.67,69.70){\line(1,0){20.00}}
\end{picture}
\caption{Two graph partitions into disjoint cycles/edges}
\label{p4}
\end{center}
\end{figure}

On Fig. \ref{p4} (a), (b) the two graph partitions into disjoint cycles/edges 
are represented that correspond to two distinct solutions of the same 
assignment problem shown on Fig. \ref{p4} (c), (d) respectively. The solutions 
of the assignment problem had been obtained as perfect matching in a bipartite 
graph \cite{christofides}. 

Thus, the following assertion is true. 

\begin{theorem}
\label{gr}
MVDCCP can not be solved by a greedy algorithm.
\end{theorem}

\begin{lemma}
\label{hmc}
If a graph $G$ be Hamiltonian then the solution of MVDCCP is a Hamiltonian
cycle. 
\end{lemma}

\noindent
{\bf Proof.} It is evident.$\circ$

\vspace{1pc}

Denote edges of the graph above: 

\begin{center}
\begin{tabular}{llll}
$e_{1}=\{x_{1}, x_{2}\}$, & $e_{2}=\{x_{1}, x_{8}\}$, & 
$e_{3}=\{x_{2}, x_{3}\}$, & $e_{4}=\{x_{2}, x_{3}\}$, \\ 
$e_{5}=\{x_{2}, x_{3}\}$, & $e_{6}=\{x_{2}, x_{3}\}$, & 
$e_{7}=\{x_{2}, x_{3}\}$, & $e_{8}=\{x_{2}, x_{3}\}$, \\ 
$e_{9}=\{x_{2}, x_{3}\}$, & $e_{10}=\{x_{2}, x_{3}\}$, & 
$e_{11}=\{x_{2}, x_{3}\}$, & $e_{12}=\{x_{2}, x_{3}\}$.\\
\end{tabular}
\end{center}

In this case, the HCP is a hereditary system $(E, Q)$, where the set $Q$ 
contains only support solution 
\begin{displaymath}
\pi^{*} = \{e_{1}, e_{2}, e_{3}, e_{7}, e_{8}, e_{9}, e_{10}, e_{12}\}.
\end{displaymath}
and all subsets of $\pi^{*}$. The set $Q$ have no other elements.

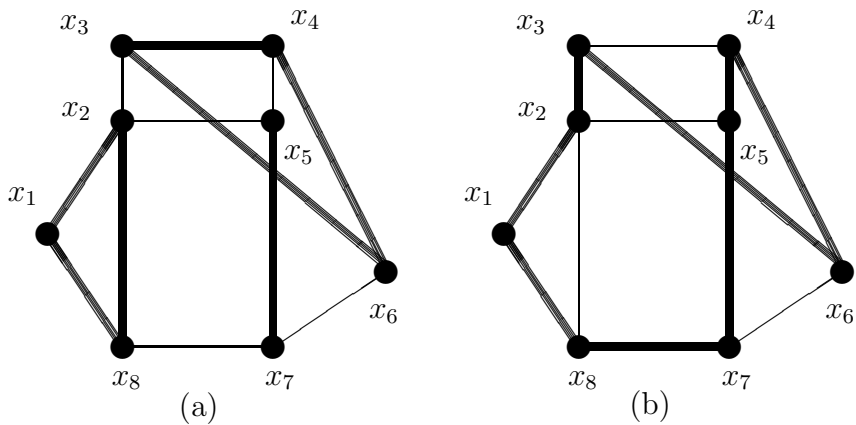
\begin{theorem}
A partial solution of HCP can not be found by a sequential method in
polynomial time.
\end{theorem}

\noindent
{\bf Proof.} At each step of a sequential method, we find a partial solution 
of a problem. Since each support solution of HCP is a Hamiltonian cycle then
at each step of the sequential method we should pick an edge of some Hamiltonian
cycle.

Let $\pi_{1}$ be some partial solution of HCP. Suppose that the next partial 
solution of HCP $\pi_{2}$, where $Card(\pi_{2}) = Card(\pi_{1})+1$, can be 
found in polynomial time. It follows that making locally optimal choice at 
each step -- an edge of a graph -- we will produce a globally optimal solution 
of HCP -- a Hamiltonian cycle, that is, we can construct a Hamiltonian cycle by 
a greedy algorithm. Hence, MVDCCP can be solved by a greedy algorithm.
It contradicts to Theorem \ref{gr}.$\circ$

\begin{corollary}
HCP does not belong to \uf.
\end{corollary}

Thus, it is proved that the Hamilton cycle problem is inherently exponential.

\begin{theorem}
\label{a3}
If a problem of the class \np\ is a hereditary system then this problem is 
effectively solvable if and only if it belongs to the class \uf.
\end{theorem}

\noindent
{\bf Proof.} By Theorem \ref{seq}, the solution of every problem $Z \in \np$ 
can be obtained by a sequential method. If the problem $Z$ is a hereditary 
system then its global solution is a support solution (see Section \ref{MOD}). 
Then the validity of Theorem \ref{a3} follows from the Theorem \ref{support}.$\circ$

\begin{theorem}
\label{reduce}
Every problem of \np\ is reduced to some problem of \uf.
\end{theorem}

\noindent
{\bf Proof.} It will suffice to indicate that MISP belongs to \uf.$\circ$

\vspace{1pc}

Of course, if the given decision problem of \np\ is reduced to another decision 
problem of \np\ then we have a {\it new} problem. The new problem may has 
{\it other} properties than the initial problem.

\end{document}